\documentclass[aps,prx,reprint,superscriptaddress]{revtex4-2}
\usepackage[utf8]{inputenc}
\usepackage{siunitx}
\usepackage{physics}
\usepackage{lipsum}
\usepackage{graphicx}
\usepackage{amsmath}
\usepackage{comment}
\usepackage{amsfonts}

\begin{document}

\title{Flexible entanglement-distribution network with an AlGaAs chip \\ for secure communications}

\author{Félicien Appas}
\affiliation{Laboratoire  Mat\'eriaux  et  Ph\'enom\`enes  Quantiques, Universit\'e de Paris, CNRS-UMR 7162, Paris 75013,  France}
\author{Florent Baboux}
\affiliation{Laboratoire  Mat\'eriaux  et  Ph\'enom\`enes  Quantiques, Universit\'e de Paris, CNRS-UMR 7162, Paris 75013,  France}
\author{Maria I. Amanti}
\affiliation{Laboratoire  Mat\'eriaux  et  Ph\'enom\`enes  Quantiques, Universit\'e de Paris, CNRS-UMR 7162, Paris 75013,  France}
\author{Aristide Lemaître}
\affiliation{Universit\'e Paris-Saclay, CNRS, Centre de Nanosciences et de Nanotechnologies, 91120, Palaiseau, France}
\author{Fabien Boitier}
\affiliation{Nokia Bell Labs, Nozay, France}
\author{Eleni Diamanti}
\affiliation{Sorbonne Universit\'e, CNRS, LIP6, 4 place Jussieu, F-75005 Paris, France}
\author{Sara Ducci}
\affiliation{Laboratoire  Mat\'eriaux  et  Ph\'enom\`enes  Quantiques, Universit\'e de Paris, CNRS-UMR 7162, Paris 75013,  France}

\date{\today}

\begin{abstract}
	Quantum communication networks enable applications ranging from highly secure communication to clock synchronization and distributed quantum computing. Miniaturized, flexible, and cost-efficient resources will be key elements for ensuring the scalability of such networks as they progress towards large-scale deployed infrastructures. Here, we bring these elements together by combining an on-chip, telecom-wavelength, broadband entangled photon source with industry-grade flexible-grid wavelength division multiplexing techniques, to demonstrate reconfigurable entanglement distribution between up to 8 users in a resource-optimized quantum network topology. As a benchmark application we use quantum key distribution, and show low error and high secret key generation rates across several frequency channels, over both symmetric and asymmetric metropolitan-distance optical fibered links and including finite-size effects. By adapting the bandwidth allocation to specific network constraints, we also illustrate the flexible networking capability of our configuration. Together with the potential of our semiconductor source for distributing secret keys over a \SI{60}{\nano\metre} bandwidth with commercial multiplexing technology, these results offer a promising route to the deployment of scalable quantum network architectures.
\end{abstract}

\maketitle

\noindent {\large \textbf{Introduction}}\\
Quantum technologies have the potential to enhance in an unprecedented way the security of communications across network infrastructures. Services and applications that large-scale quantum communication networks can provide span secure communication with security guarantees impossible to achieve only with classical systems~\cite{Diamanti2016}, delegated and blind quantum computing~\cite{Gheorghiu2019}, or distributed quantum computing and sensing~\cite{Zhao2020}, eventually leading to the full capabilities of a quantum internet~\cite{Wehner2018}. The choice of network topology and the optimization of the corresponding required resources are crucial when designing the architecture of such infrastructures. This is particularly true for the entangled photonic resources that need to be deployed to enable some of the most advanced applications, but also to alleviate the need for trusted nodes in cryptographic scenarios~\cite{Yin2020,Diamanti2020}.

In this context, a topology that has naturally attracted much attention exploits the presence of polarization and time-energy photonic entanglement in state-of-the-art sources to distribute polarization-entangled photons from a single source to multiple receivers (users) that hold energy-matched channel pairs~\cite{Lim2008,Herbauts2013,Aktas2016,Autebert2016,Wengerowsky2018,Zhu2019,Joshi2020,Lingaraju2020}. This concept was in particular ingeniously pushed in~\cite{Wengerowsky2018}, which proposed and demonstrated a scheme allowing each user of the network to share quantum correlations with every other user, hence forming a fully connected entanglement network. Several challenges come, however, with such an architecture. First, the number of required frequency channels increases quadratically with the number of users, making the bandwidth of the source a critical feature for ensuring scalability to large multi-user networks. Although the scaling was improved in~\cite{Joshi2020}, the relatively narrowband polarization entangled photon pair source based on periodically poled crystals that was used would not permit to further increase the number of users. Furthermore, scalability can also be hindered by the use of cascaded passive elements such as dense wavelength division multiplexers (DWDM) and beamsplitters,  which also come with high losses~\cite{Wengerowsky2018,Joshi2020}. Employing instead advanced multiplexing techniques based on wavelength selective switch (WSS) technology offers a number of attractive features for flexible entanglement distribution, as was recently shown in~\cite{Lingaraju2020}, albeit again with a relatively narrowband entangled photon source.

In this work, we demonstrate a scalable approach to the fully connected architecture of~\cite{Wengerowsky2018}. To this end, we use a broadband source, emitting photon pairs at telecom wavelength that is based on an AlGaAs semiconductor chip. AlGaAs exhibits direct bandgap, strong electro-optical effect, and small birefringence, which enables an effective miniaturization and the generation of polarization-entangled states without any off-chip compensation~\cite{Orieux2013,Horn2012,Kang2016}. We show that thanks to the features of our source, we can maintain an entanglement fidelity above 85\% over a \SI{60}{\nano\metre}-wide spectral region and across long-distance fiber links. To benchmark the performance of our system, we then run the flagship quantum communication application, namely QKD, and in particular the BBM92 protocol~\cite{Bennett1992}, both in the asymptotic and finite-size regime, over symmetric and asymmetric fibre-optic links, separating the users by up to 50~km in the symmetric case. We distribute entanglement using Flexgrid multiplexing based on WSS technology, which is the current industry standard in classical communications, and offers straightforward reconfigurability in terms of central frequency and channel bandwidth adjustment, wavelength-insensitive insertion loss and polarization diversity. This allows us to show that each of the two-user links in a fully connected entangled network of 4, 5 or 8 users, can support high-quality entanglement distribution and efficient QKD at metropolitan scale distances. We further showcase the flexibility of our scheme in an unbalanced network scenario, featuring several local users and one distant user. We demonstrate that the bandwidth reallocation enabled by the WSS can be used to equilibrate the resulting rates, hence materializing an elastic network configuration and highlighting its potential for the deployment of flexible, scalable quantum network architectures.\\

\begin{figure*}
	\includegraphics[width=\textwidth, keepaspectratio]{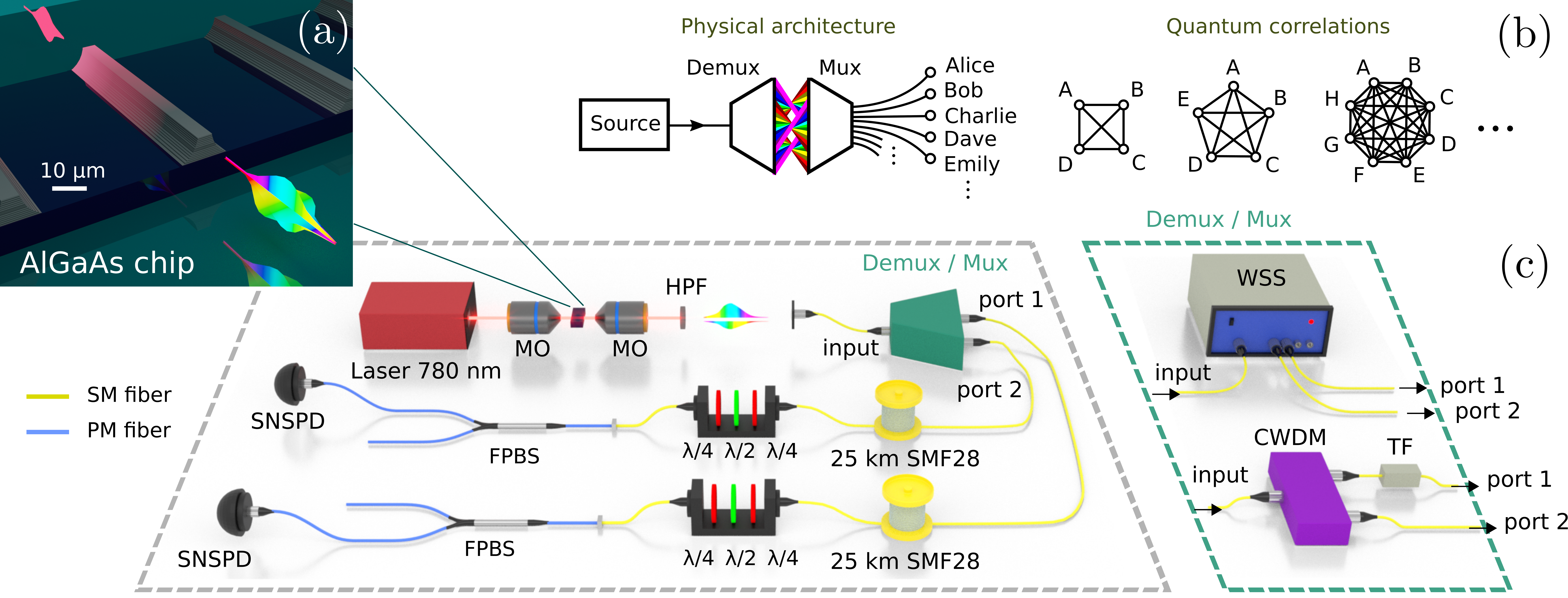}
	\caption{\label{fig:setup} (a) Artist view of an AlGaAs chip under operation. An incoming pump photon is depicted on the top left corner, at the input of the waveguide. The generated entangled two-photon state is represented on the bottom right corner. (b) Schematic view of the entanglement network architecture. The physical structure of the network consists of an entangled photon source, a demux/mux stage and a distribution stage. The resulting quantum correlation topology takes the form of a fully-connected graph of variable size (here $N=4,5, 8$), where each user node shares an entangled state with all other nodes in the network. (c) Experimental setup. MO: Microscope Objective. HPF: High Pass Filter. WSS: Wavelength Selective Switch. CWDM: Coarse Wavelength Division Multiplexing unit. TF: Tunable Filter. FPBS: Fibered Polarizing Beam Splitter. SNSPD: Superconducting Nanowire Single Photon Detector. SM: Single Mode fiber. PM: Polarization Maintaining fiber.}
\end{figure*}

\noindent {\large \textbf{Results}}\\
\noindent\textbf{AlGaAs photonic chip and experimental setup.}
The schematic of our experimental setup is presented in Fig.\:\ref{fig:setup}\:(c) and consists of three stages: entanglement generation with an AlGaAs chip, frequency demultiplexing/multiplexing (demux/mux) of the generated signal, and a distribution stage. Note that two different frequency demultiplexing/multiplexing approaches can be implemented, depending on the spectral region of interest: in the telecom C-Band (\SIrange{1530}{1565}{\nano\metre}) we use a WSS and in the L-Band (\SIrange{1565}{1625}{\nano\metre}) a coarse wavelength division multiplexing unit (CWDM) with \SI{13}{nm} wide channels followed by a tunable filter. The WSS enables the implementation of flexible reconfigurable fully-connected multi-user entanglement networks as shown in Fig.\:\ref{fig:setup}\:(b).

The AlGaAs chip, depicted in Fig.\:\ref{fig:setup}\:(a), consists of a Bragg reflection ridge waveguide emitting photon pairs through type II spontaneous parametric down conversion (SPDC) in the C+L-Band \cite{Autebert2016}. The sample is pumped with a tunable CW diode laser leading to a pair production rate of the order of \num{e7} pairs/s at the chip output, corresponding to a brightness of \num{3.6e5} pairs/s/mW (see Methods section for details).

Thanks to the dispersion and nonlinear properties of the AlGaAs platform, the quantum state produced by the chip presents several features making it appealing for the implementation of quantum networks. Our chip generates time-energy and polarization entangled photons emitted in a $\ket{\Psi^{+}}$ polarization Bell state~\cite{Orieux2013,Autebert2016}. The pairs being generated in a broad spectral band, we can multiplex the generated photons in a large number of wavelength channels using standard telecom components and share polarization entanglement between users receiving energy-matched channels \cite{Autebert2016}. Besides, the group velocity mismatch between orthogonally polarized photons in the AlGaAs chip is so small that no off-chip walk-off compensation is required to obtain polarization entanglement \cite{Horn2012,Orieux2013}. This is a key feature enabling the direct use of the emitted pairs at the output of the chip, opening the way to its easy integration into simple and robust architectures.

\begin{figure}
	\includegraphics[width=0.45\textwidth, keepaspectratio]{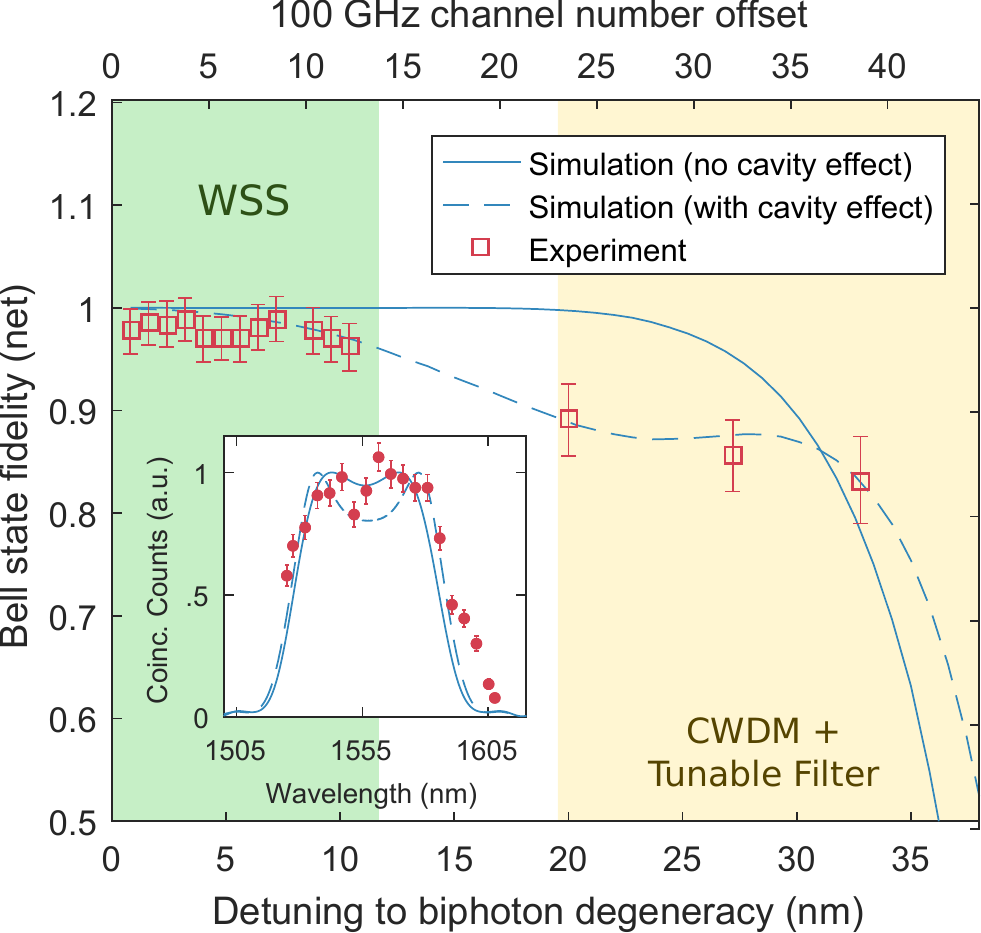}
	\caption{\label{fig:fidelity} Main graph: Lower bound to the fidelity to a $\ket{\Psi^{+}}$ Bell state as a function of the detuning to biphoton degeneracy wavelength (\SI{1556.55}{\nano\metre}) in nm (lower x-axis) and in units of 100 GHz channels (upper x-axis). Error bars take into account the Poissonian statistics of both signal counts and subtracted background counts. The different color areas correspond to the two different wavelength demultiplexing schemes (see main text for details). Inset: Coincidence counts as a function of filter wavelength.}
\end{figure}

The biphoton bandwidth has been measured by injecting the generated photon pairs into a fibered 50/50 beam splitter (BS) with a fibered C+L-Bands tunable filter inserted in one of its output ports and detecting the coincidences between the two output ports via superconducting nanowire single photon detectors (SNSPD) and a time-to-digital converter (TDC). The result is shown in the inset of Fig.\:\ref{fig:fidelity}: the full-width at half-maximum of the output signal is \SI{60}{\nano\metre}, corresponding to approximately 72 ITU  \SI{100}{\giga\hertz} channels, 36 on each side of the quantum state degeneracy frequency.\\


\noindent\textbf{Broadband entanglement characterization.}
In order to assess the effective biphoton bandwidth for entanglement distribution we have selected symmetric output channels with respect to the biphoton degeneracy frequency and we have estimated the fidelity of the transmitted photon pairs to the $\ket{\Psi^{+}}$ Bell state as a function of the channel number.
In the spectral region around degeneracy (C-Band) the measurement has been performed by defining channels using the WSS, while in the spectral region far from degeneracy (L-Band) we have used the CWDM followed by a fibered tunable filter. As shown in Fig.\:\ref{fig:setup}\:(c), after demultiplexing, the photons are sent through a polarization analysis stage consisting of a set of $\lambda/4,\lambda/2,\lambda/4$ waveplates, then through a fibered polarizing beam splitter (FPBS), and finally detected. The waveplates and FPBS are used to project the polarization of the measured photons onto the states of two mutually unbiased bases (MUB), namely  $\mathbb{X} = \{\ket{H}, \ket{V}\}$ and $\mathbb{Z} =\{\ket{D} = \frac{1}{\sqrt{2}}(\ket{H} + \ket{V}), \ket{A} = \frac{1}{\sqrt{2}}(\ket{H} - \ket{V})\}$. A lower bound to the fidelity $F$ to the $\ket{\Psi^{+}}$ Bell state is estimated by coincidence measurements in those two bases by recording the number of coincidence counts between the two detectors in 8 different configurations~: $C_{HH}, C_{HV}, C_{VH}, C_{VV}, C_{DD}, C_{DA}, C_{AD}, C_{AA}$ (see Methods for details). From these quantities, we can access the diagonal elements of the density matrix in both MUBs. For $(\alpha,\beta) \in \mathcal{B}\times\mathcal{B}$, with $\mathcal{B} \in \lbrace \mathbb{X}, \mathbb{Z} \rbrace$, the matrix element $\rho_{\alpha\beta}$ corresponding to the first (second) photon in $\alpha\:(\beta)$ polarization state is calculated as
$\rho_{\alpha\beta} = C_{\alpha\beta}/(\sum_{\mathcal{B}\times\mathcal{B}}C_{\alpha'\beta'})$. Note that the normalization factor has to be calculated independently for the two bases.
The lower bound to $F$ is then obtained through the relation~\cite{Chang2016,Blinov2004}:
\begin{align}\label{eq:fidelity_bound}
F \ge\ & \frac{1}{2}\ (\rho_{HV} + \rho_{VH} - \rho_{HH} - \rho_{VV} \\
& + \rho_{DD} + \rho_{AA} - \rho_{DA} - \rho_{AD}). \nonumber
\end{align}

The experimental results, along with the fidelity of the numerically simulated quantum state emitted by the AlGaAs chip, are shown in Fig.\:\ref{fig:fidelity}. The observed lower bound of the fidelity is above \SI{95}{\percent} over a \SI{26}{\nano\metre} spectral range corresponding to the first 13 pairs of \SI{100}{\giga\hertz} ITU channels around degeneracy, and stays above \SI{85}{\percent} for up to 38 channels pairs around degeneracy, spanning a total \SI{60}{\nano\metre} wavelength range. Since the channels of the CWDM are not centered around the biphoton degeneracy wavelength, measurements in the L-Band are limited to the conjugate channels that fall within the transmission window of the CWDM ports. For this reason only 3 data points have been acquired in the orange region of Fig.\:\ref{fig:fidelity}.

Close to degeneracy, the experimental data are in good agreement with theoretical predictions describing the SPDC process without taking into account facets reflectivity of the AlGaAs waveguide (solid lines). However, moving away from degeneracy, cavity effects arising from facets reflectivity have to be taken into account. The inclusion of these effects in our theoretical model leads to the calculated fidelity represented with a dashed line, reproducing the experimental data with a very good quantitative agreement. Details on the calculation of the fidelity can be found in the Methods.

The ability to maintain high quality polarization entanglement over a broad spectral region establishes the high potential of AlGaAs-based devices as ready-to-use miniature sources of broadband polarization entangled biphoton states.\\	
	

\noindent\textbf{QKD performance before the long-distance distribution stage.}
Next, we exploit the broadband polarization entangled biphoton state generated with the AlGaAs chip to implement the BBM92 QKD protocol \cite{Bennett1992}.
In this scenario, two users share a maximally entangled Bell state, in our case: $\ket{\Psi^{+}} = \frac{1}{\sqrt{2}}\left( \ket{H}_{s}\ket{V}_{i} + \ket{V}_{s}\ket{H}_{i}\right)$ with signal $(s)$ and idler $(i)$ denoting the high and low energy photon respectively. We deterministically separate the photons of each pair into two fiber paths according to their frequency ($s$ or $i$) using the WSS. As explained previously, each photon is then sent to a measurement station, consisting of a polarization analysis stage and a SNSPD as shown in Fig.\:\ref{fig:setup}\:(c).
In the full BBM92 protocol, the basis choice is random, each user having 4 detectors, one for each measurement outcome ($H,V,D,A$). The random basis choice can be implemented by using, for instance, passive 50/50 BS \cite{Ursin2007} or polarization-to-time-bin conversion \cite{Zhu2019}. However, in our proof-of-principle measurement, we choose for simplicity to manually configure the waveplates in each path ($s$ and $i$) and record the coincidence counts in the same 8 configurations as presented previously, in order to assess the performance of our QKD scheme.



To evaluate the intrinsic QKD performance of our scheme, we measure the figures of merit of the BBM92 protocol right after the demultiplexing stage:
a) the asymptotic secret key rate $R_{\mathrm{key}}$, which corresponds to the number of secret bits per second established by the two users following basis reconciliation (sifting), error correction and privacy amplification, and b) the quantum bit error rate $e$ (QBER), which is the fraction of erroneous bits in the raw key.
We repeat the measurement for 13 different choices of signal and idler \SI{100}{\giga\hertz} channel pairs.
Note that this spectral range is limited by the upper cutoff wavelength (\SI{1565}{\nano\metre}) of the WSS corresponding to the boundary of the C-Band, and not by the spectral bandwidth of the generated biphoton state.

We first express $R_{\mathrm{key}}$ and $e$ in terms of the recorded coincidence counts \cite{Zhu2019}. The raw coincidence counts in each basis ($\mathbb{X}$ and $\mathbb{Z}$) read:
\begin{align}\label{eq:rawcounts}
C_{\mathbb{X}}^{(\mathrm{raw})} &= C_{HH} + C_{HV} + C_{VH} + C_{VV} \\
C_{\mathbb{Z}}^{(\mathrm{raw})} &= C_{DD} + C_{DA} + C_{AD} + C_{AA}
\end{align}
The QBER can be readily obtained by taking the ratio of accidental counts against the total number of recorded raw coincidence counts:
\begin{equation}\label{eq:qber}
e = \dfrac{C_{HH} + C_{VV} + C_{DA} + C_{AD}}{C_{\mathbb{X}}^{(\mathrm{raw})} + C_{\mathbb{Z}}^{(\mathrm{raw})}}.
\end{equation}
Next, we compute the raw key rate as the arithmetic mean between the raw count rate in the $\mathbb{X}$ and in the $\mathbb{Z}$ basis:
\begin{equation}\label{eq:rraw}
R_{\mathrm{raw}} = \dfrac{1}{2}\dfrac{C_{\mathbb{X}}^{(\mathrm{raw})} + C_{\mathbb{Z}}^{(\mathrm{raw})}}{\tau},	
\end{equation}
with $\tau$ the integration time.
Finally, the asymptotic key rate is obtained as \cite{Luetkenhaus2000,Ma2007}:
\begin{equation}\label{eq:asympt_keyrate}
R_\mathrm{key} \ge R_\mathrm{raw} \dfrac{1}{2} (1-f(e)H_2(e)-H_2(e)),
\end{equation}
where $H_2(e) = -e\log_2(e)-(1-e)\log_2(1-e)$ is the binary entropy function. This corresponds to the maximum number of key bits per unit time that can be securely extracted by the two parties in the limit of infinite key length. The factor $1/2$ accounts for the basis reconciliation process (sifting) of the BBM92 protocol.  $f(e)H_2(e)$ is the portion of the key that has to be used for error correction, with $f(e) \ge 1$ a function which quantifies the efficiency of the error-correction code. In this work, we use an interpolation of tabulated values for $f(e)$ \cite{Luetkenhaus2000} assuming the standard bidirectionnal code presented in Ref.\:\cite{Brassard1994}.
Finally, the last term in $H_2(e)$ is the lower bound of the fraction of the raw key that is lost after privacy amplification \cite{Bennett1995}.

\label{sec:qkd_0km}
\begin{figure}
	\includegraphics[width=0.45\textwidth, keepaspectratio]{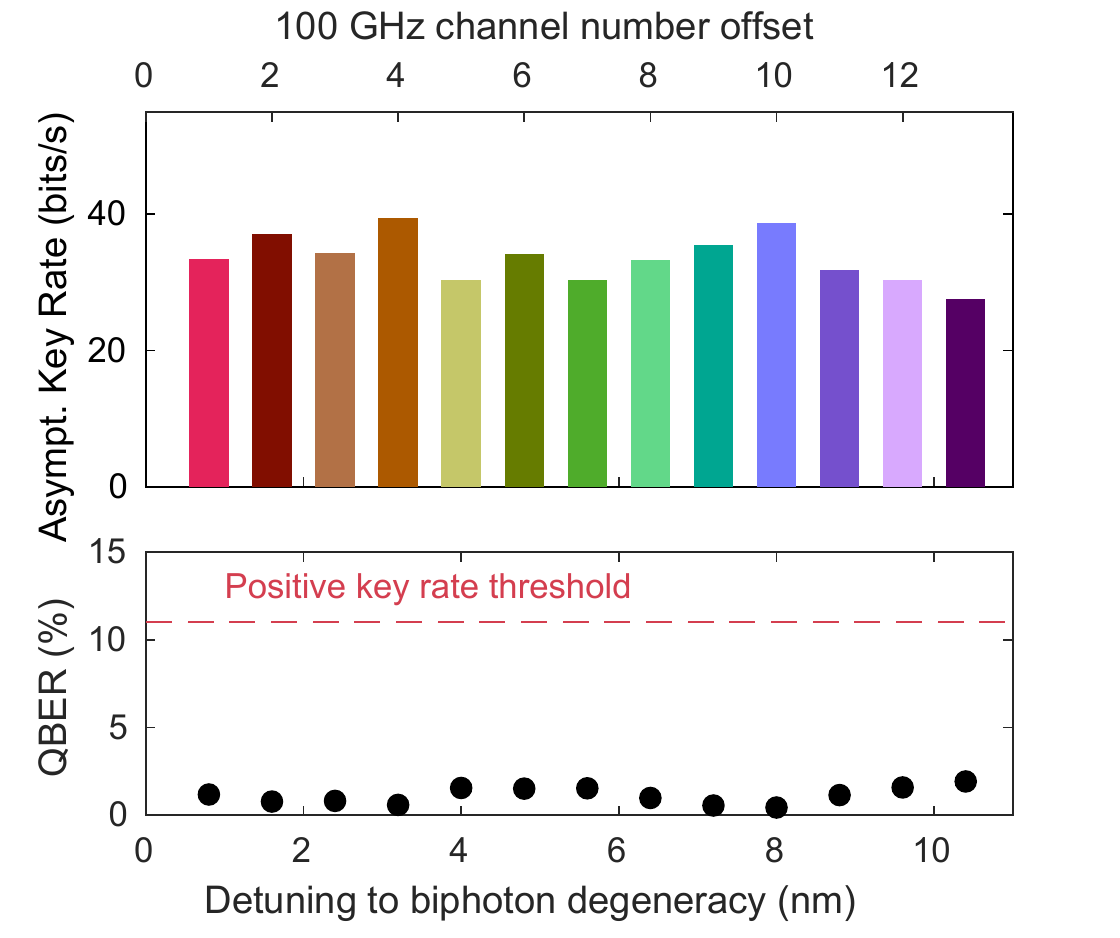}
	\caption{\label{fig:keyrate_qber_wl} Asymptotic BBM92 key rate (upper panel) and QBER (lower panel) as a function of the detuning to biphoton degeneracy in nm (lower x-axis) and in units of 100 GHz channels (upper x-axis). The \SI{11}{\percent} error threshold for a positive key rate is given for an error correcting code operating at the Shannon limit.}
\end{figure}

The calculated $R_\mathrm{key}$ and QBER are plotted in Fig.\:\ref{fig:keyrate_qber_wl}. We see that both quantities are stable over the 13 ITU \SI{100}{\giga\hertz} channels spanning a \SI{10.4}{\nano\metre} wavelength range. The QBER stays below \SI{2}{\percent}, well under the positive key rate threshold of \SI{11}{\percent} for which $1-f(e)H_2(e)-H_2(e)=0$, assuming an error-correcting code operating at the Shannon limit ($f(e)=1$). This stability can be attributed to the flatness of the source spectrum and fidelity in this spectral region (Fig.\:\ref{fig:fidelity}) and to the wavelength-independent insertion losses of the WSS.
The entanglement quality of the emitted quantum state, combined with the high conversion efficiency of AlGaAs, yields high asymptotic key rates of 28 to 39 bits/s, over a very broad spectral range.
The ability to support high key rates with low error over many frequency channels makes our AlGaAs entangled-photon source an ideal technology for a chip-based quantum network, as will be illustrated later.\\

\begin{figure*}
	\includegraphics[width=\textwidth, keepaspectratio]{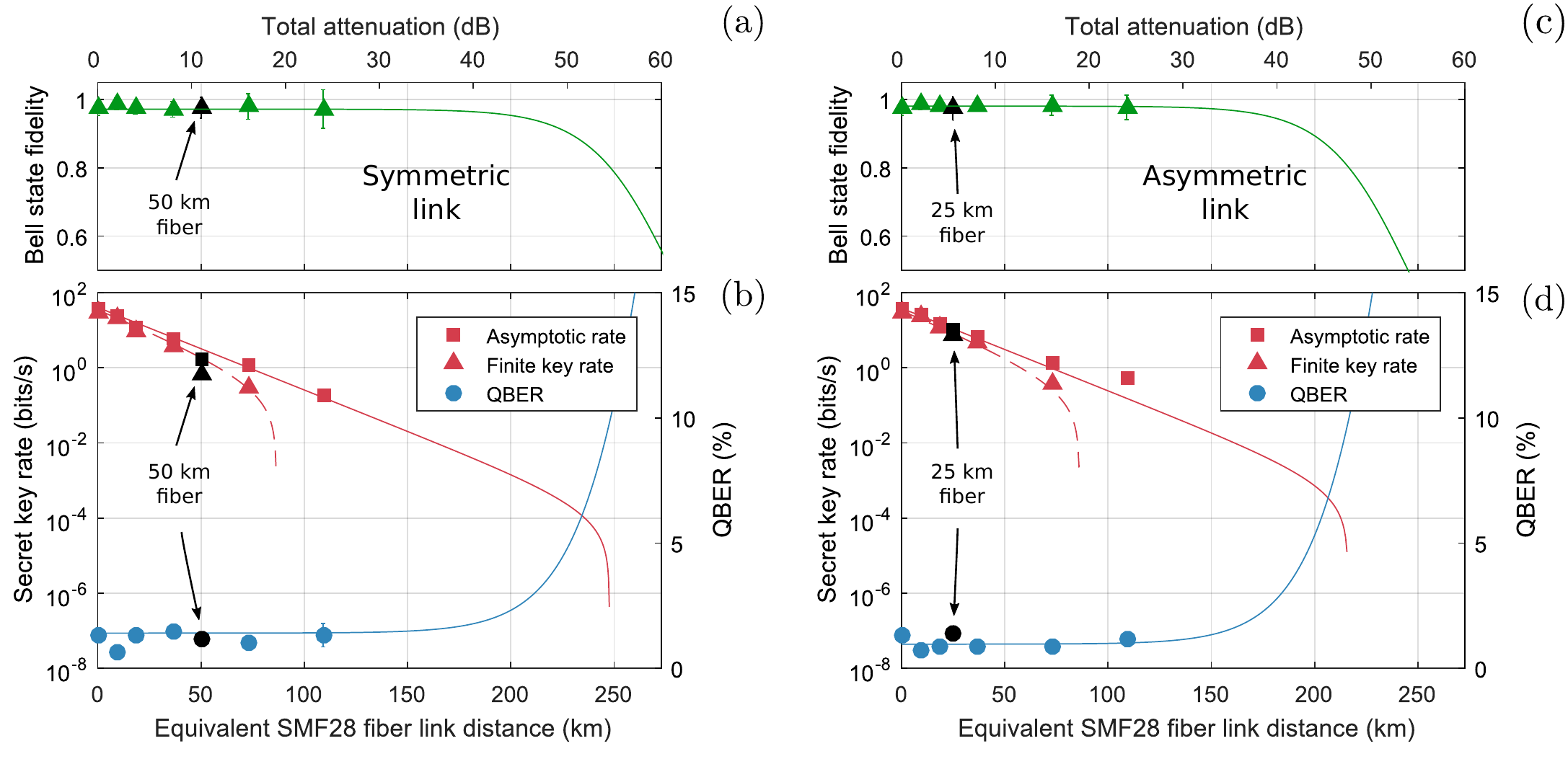}
	\caption{\label{fig:longdist} Lower bound to the fidelity to a $\ket{\Psi^{+}}$ Bell state, asymptotic and finite size key rates and QBER for a (a,b) symmetric and an (c,d) asymmetric two-user link supported by 100 GHz ITU channels 23 (\SI{1558.98}{\nano\metre}) and 29 (\SI{1554.13}{\nano\metre}) as a function of SMF28 fiber distance (lower x-axis) and attenuation (upper x-axis). Symbols represent experimental data and continuous lines theoretical predictions for a BBM92 scheme. Finite key rates are estimated assuming \SI{10}{\minute} block size.}
\end{figure*}

\noindent\textbf{Entanglement distribution and secret key distribution over long-distance fiber links.}
After quantifying the intrinsic entanglement generation and QKD performance of the source over a broad spectral range, we assess the long-distance performance of our architecture.
To certify the capability of our setup to maintain high-quality entanglement over long fiber links, we measure the lower bound to the fidelity given in Eq.~\eqref{eq:fidelity_bound} after adding \SI{25}{\kilo\meter} SMF28 fiber spools between the demux/mux stage and the user's measurement station. This measurement is performed in two configurations: symmetric (both users separated from the source by \SI{25}{\kilo\meter} of fiber) and asymmetric (one user at \SI{0}{\kilo\meter} and the other at \SI{25}{\kilo\meter}) in order to compare the performances and provide the optimized solution for entanglement distribution. All measurement runs are performed by selecting entangled photons from \SI{100}{\giga\hertz} ITU channels 23 (\SI{1558.98}{\nano\meter}) and 29 (\SI{1554.13}{\nano\meter}) chosen as representative of all the available channels. The results are shown as black symbols on Fig.\:\ref{fig:longdist}\:(a) and (c) for the symmetric and asymmetric configurations respectively.

To estimate the maximum distance of a repeaterless link we also perform this measurement by replacing the fiber spools by variable channel attenuation programmed on the WSS. The attenuation can then be converted into equivalent length of SMF28 fibers assuming a standard value of \SI{0.22}{\decibel\per\kilo\meter} for the SMF28 optical losses. Results for a symmetric and asymmetric link are shown in Fig.\:\ref{fig:longdist}\:(a) and (c) respectively as color symbols.
The theoretical value of the lower bound to $F$ as a function of channel attenuation (solid lines in Fig.\:\ref{fig:longdist}\:(a) and (c)) is computed using the model of Ref.\:\cite{Ma2007}.

We observe that the fidelity stays above \SI{95}{\percent} up to \SI{204}{\kilo\metre} in the symmetric case and \SI{178}{\kilo\metre} in the asymmetric case. At longer distances, the decrease of the signal-to-noise ratio induces a drop of the estimated fidelity. We conclude that our entanglement distribution scheme is robust to link asymmetry enabling applications in real-world networks of arbitrary topology.

We stress that the performance of our device being practically insensitive to wavelength in the C-band, as shown in Fig.\:\ref{fig:fidelity}, the measured performance for the two particular channels (23 and 29) can be extrapolated to the 12 other available \SI{100}{\giga\hertz} channel pairs within the operating range of the WSS.
This means that each channel pair can support a high-quality long-distance entanglement distribution channel, making our scheme compatible with entanglement-based quantum information protocols over metropolitan networks.


As a benchmark application, we run once again the BBM92 protocol onto our system and evaluate its performance. To do so, we estimate $R_{\mathrm{key}}$ and $e$ in the presence of fiber spools (black symbols in Fig.\:\ref{fig:longdist}\:(b) and (d)) and variable attenuation in both symmetric and asymmetric configurations (color symbols in Fig.\:\ref{fig:longdist}\:(b) and (d)).
We compare the experimental values of the asymptotic key rate to theoretical predictions (continuous lines) for a BBM92 scheme with passive basis selection following the model of Refs.\:\cite{Scheidl2009, Ma2007}.

Furthermore, in order to highlight the relevance of these results for real-world implementations, we perform a finite key security analysis of our scheme.
As mentioned previously, the secret key rate in Eq.~\eqref{eq:asympt_keyrate} is only valid in the limit of infinite key length.
However, in practice, raw key bits are exchanged in blocks of finite size. To guarantee a certain security threshold for a given block size, we resort to the framework developed in Refs.\:\cite{Tomamichel2012,Yin2020}. We set correctness and security bounds to $\epsilon_\text{corr} = \num{e-10}$ and $\epsilon_\text{sec} = \num{e-10}$ respectively for a \SI{10}{\minute} block size. The resulting key rates are shown in Fig.\:\ref{fig:longdist} (b) and (d) as triangles for experimental values and dashed lines for theoretical predictions.


Based on this data, we can compare experimentally the long-distance performance of the BBM92 protocol in symmetric and asymmetric configurations.
We observe that, in both scenarios, the key rate including finite size effects stays positive for distances of up to \SI{75}{\kilo\meter}. However, we also observe that, at very long distances, the asymptotic key rate for a symmetric link stays positive up to \SI{250}{\kilo\metre}, while in an asymmetric link, it drops at around \SI{215}{\kilo\metre}. Indeed, in the latter configuration, a strong attenuation in the link between the source and the distant user induces a strong increase of the QBER as the signal approaches the noise background. On the contrary, in the symmetric case, the losses are distributed between the two users and this critical situation is reached at higher levels of attenuation. In the finite key security regime in which we are operating, the finite key rates become negative before the difference between symmetric and asymmetric links comes into play. This result has practical implications in the context of deployed QKD schemes; indeed this proves that a scenario involving an entangled pair source connected to one local user and one distant user will not be detrimental to the BBM92 protocol efficiency, hence enabling the use of our scheme in realistic configurations.\\


\noindent\textbf{Multi-user entanglement networks with flexible bandwidth allocation.}
Building on the previously presented results, we now demonstrate the operation of a scalable multi-user network architecture by taking advantage of the broad entanglement bandwidth of our source and of the flexible frequency management enabled by the WSS.

We consider a simple physical network consisting of one central server node which hosts the entangled photon source, and $N$ user nodes. Each user has a polarization analyzing device and single photon detectors to carry out the BBM92 protocol. We focus on a particular class of such networks, where each of the $N$ users shares an entangled state with the remaining $N-1$ users \cite{Wengerowsky2018}. We refer to these as fully-connected networks, by analogy with graph theory: indeed, they can be represented by a complete graph where each vertex corresponds to one user node and the edges symbolize a shared entangled state between two users. Graph representations for fully-connected entanglement networks of size $N=4,5,8$ are shown in Fig.\:\ref{fig:setup}\:(b).
The number of edges in a complete graph of size $N$ is $N(N-1)/2$. Hence to get a complete entanglement network, one needs to establish $N(N-1)/2$ two-user links supported by $N(N-1)/2$ distinct conjugate channel pairs. To do so, we demultiplex the generated photons into $N(N-1)$ frequency channels. Then we recombine those channels into single optical fibers, one for each user. This is done via the WSS which implements both operations. Each user then receives photons from $N-1$ channels, one for each connected node. Photons from the $N-1$ conjugate channels are similarly distributed to the other users. As a result, each node effectively shares a Bell pair with every other node in the network.
The number of required frequency channels scales quadratically with $N$. In this context, we see that the broadband entanglement of the biphoton state emitted by the AlGaAs chip is a crucial asset.
In addition, the use of a WSS at the demultiplexing/multiplexing stage enables a reconfigurable multi-user network where the central frequency and the bandwidth of each channel can be adjusted. In our setup we can increase the number of users in the network by simply reducing the bandwidth allocated to each channel. Using all the available wavelength overlap between the biphoton state and the WSS operating range, our reconfigurable complete-graph network can accommodate from 4 users, with \SI{200}{\giga\hertz} channels, to 8 users, where each link is supported by a \SI{50}{\giga\hertz} conjugate channel pair.

\begin{figure*}
	\includegraphics[width=\textwidth, keepaspectratio]{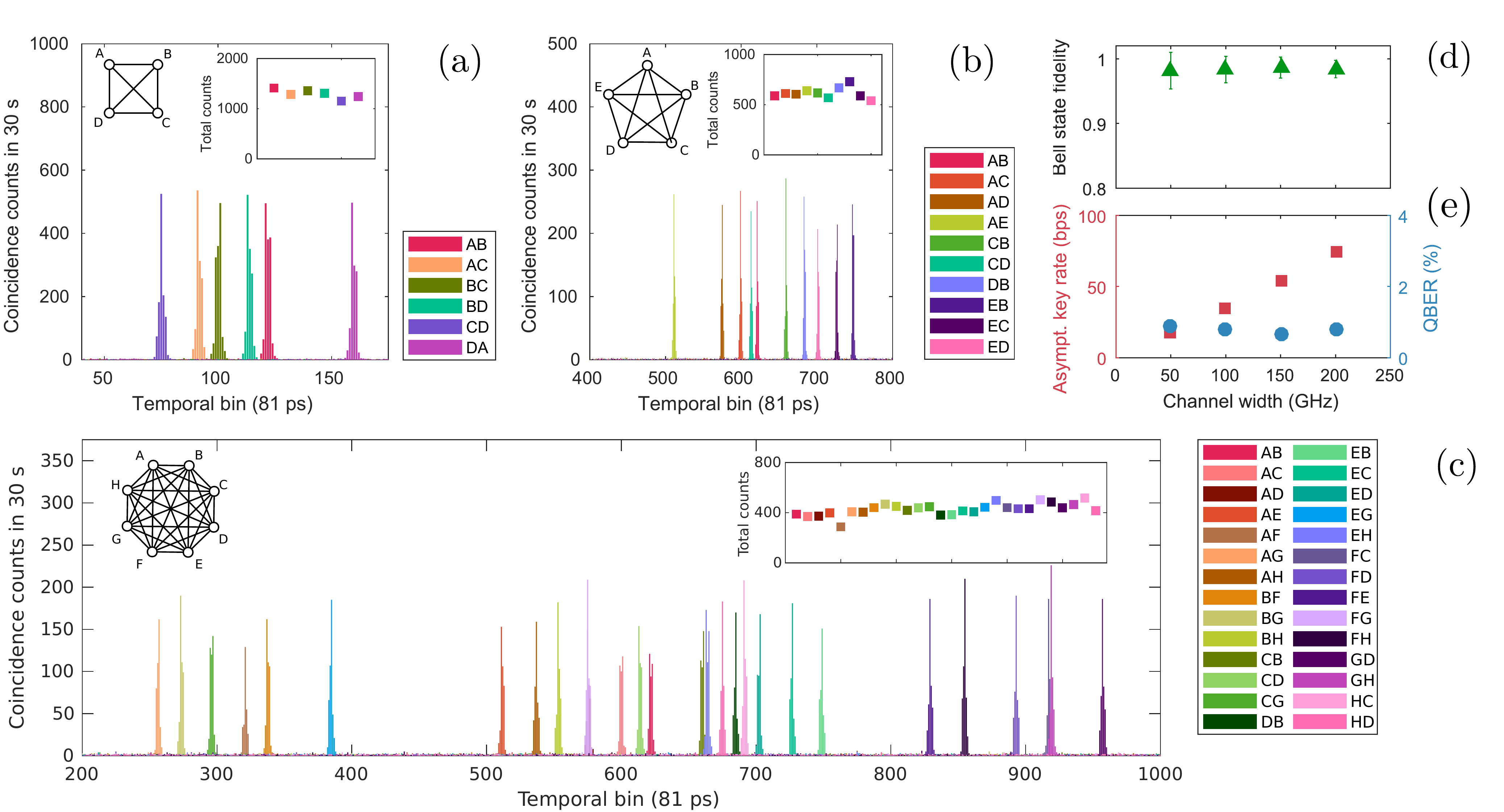}
	\caption{\label{fig:multiuser} Reconfigurable multi-user entanglement network. (a-c): Time-correlation histograms for all two-user links in a fully connected network of size (a) $N=4$ with 200 GHz channels, (b) $N=5$ with 100 GHz channels and (c) $N=8$ with 50 GHz channels. (d) Lower bound to the fidelity to a $\ket{\Psi^{+}}$ Bell state and (e) Measured asymptotic key rate and QBER for ITU channels 23 (\SI{1558.98}{\nano\metre}) and 29 (\SI{1554.13}{\nano\metre}), as a function of channel width.}
\end{figure*}
We sequentially characterize the $N(N-1)/2$ two-user links for different network sizes: $N=4, N=5$ and $N=8$.
Due to the lack of additional single photon detectors, we did not run real-time QKD sessions for $N$ users in parallel. Instead, we measure the time-correlation histograms for all $N(N-1)/2$ two-user links in the three different configurations. The obtained coincidence histograms are shown in Fig.\:\ref{fig:multiuser}\:(a-c). The total number of counts contained in each coincidence peak is reported in the insets. We observe that the number of recorded coincidence  counts is stable across all links with a relative standard deviation of \SI{7}{\percent}, \SI{9}{\percent} and \SI{11}{\percent} for $N=4,5,8$ respectively. As already discussed previously, this stability is a consequence of the broadband character of the source and of the wavelength-insensitive insertion losses of the WSS. In contrast, in networks based on passive thin-film WDM filters, the channel transmission can be subject to strong variations, leading to unbalanced networks where some links carry more signal than others.

We consolidate this result by recalling that, as shown in Fig.\:\ref{fig:fidelity} and Fig.\:\ref{fig:keyrate_qber_wl}, the distributed quantum state fidelity and BBM92 asymptotic key rate and QBER have a very weak dependence on the frequency channel. Therefore, in view of QKD applications, we expect the key rate to be evenly distributed among all two-user links in each configuration.

We further check that the entanglement quality is not reduced when changing the channel bandwidth. To do so, we estimate the lower bound to $F$ for ITU channels 23 and 29 as a function of channel bandwidth. The measurement result, displayed in Fig.\:\ref{fig:multiuser}\:(d), shows that the fidelity is essentially insensitive to the channel bandwidth. Accordingly, we observe that the QBER has also a very weak dependence on the bandwidth and the asymptotic key rate features a linear scaling, as shown in  Fig.\:\ref{fig:multiuser}\:(e).
We conclude that these values for $R_\mathrm{key}$ and QBER, alongside with those of Fig.\:\ref{fig:keyrate_qber_wl}, provide an estimate for the achievable QKD performances in the various $N$-user graphs presented in Fig.\:\ref{fig:multiuser}\:(a-c).

Our scheme presents several advantages in terms of scalability over passive WDM quantum networking schemes. 
Indeed, the flexibility offered by the WSS makes it possible to fully reconfigure the network without modifying the hardware. On the contrary, using passive WDM demux/mux the addition of more users to the network requires either changing the complex combination of cascaded WDM filters or non-deterministically splitting some channels by combining WDM filters and multiport fiber splitters \cite{Joshi2020, Liu2020}, which, either way, comes at the cost of extra optical losses. In contrast, the fixed insertion losses of the WSS makes it possible to extend the network without degrading the signal, a major asset in a fully deployed network scenario.\\

\begin{figure}
	\includegraphics[width=0.45\textwidth, keepaspectratio]{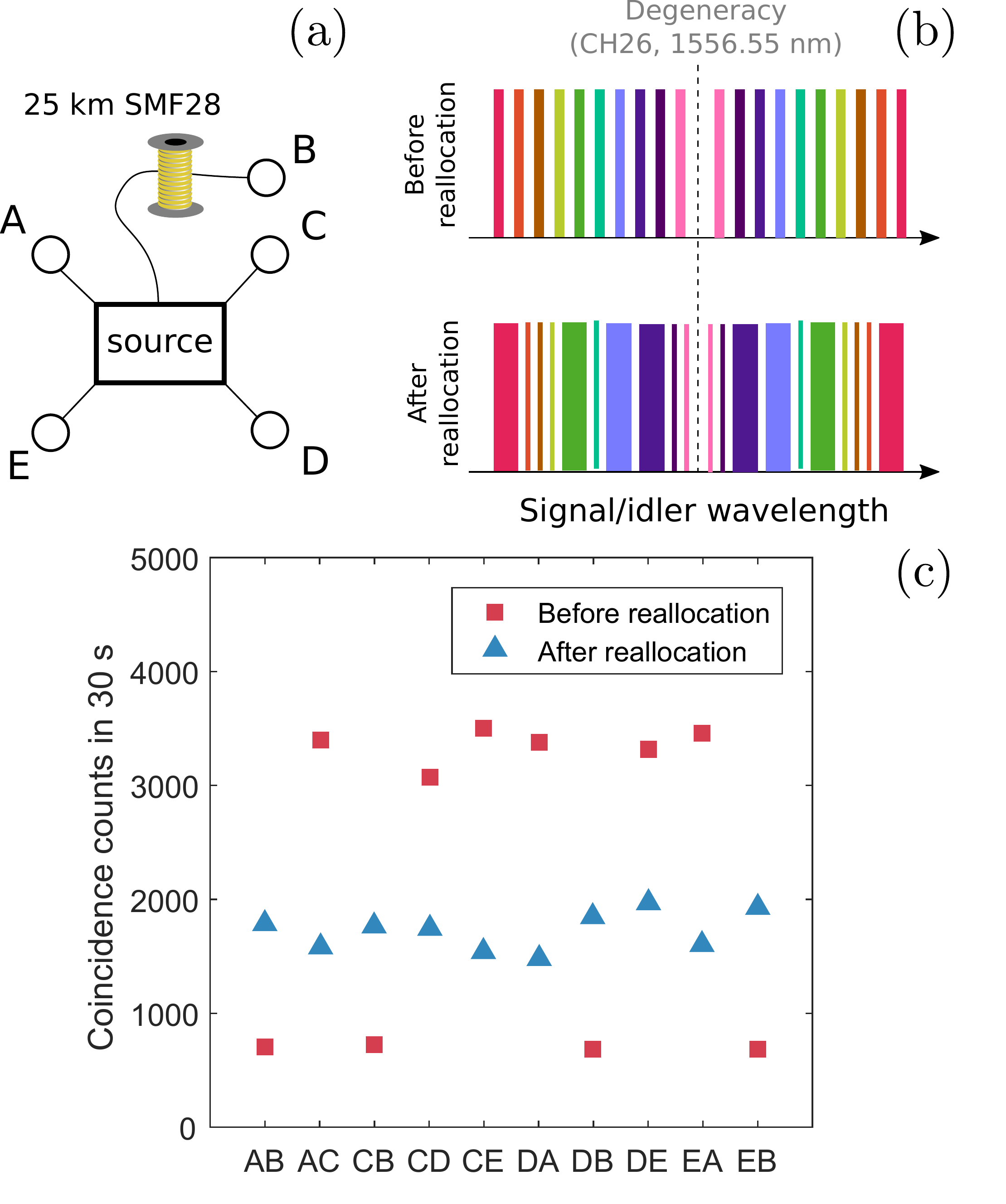}
	\caption{\label{fig:flexgrid} Illustration of flexible bandwidth allocation. (a) Sketch of the experiment. In a fully connected network with $N=5$ users, user B is separated from the source by a \SI{25}{\kilo\metre} long SMF28 fiber spool. (b) Schematic of the bandwidth distribution between all two-user links before and after applying the bandwidth reallocation algorithm. After reallocation, the four links (red, dark green, light blue and dark blue) that connect user B to the rest of the network are allowed a broader bandwidth. (c) Coincidence counts for each two-user link before (squares) and after (triangles) bandwidth reallocation.}
\end{figure}
Finally, we exploit the flexibility of our scheme to demonstrate a quantum network where the signal of each link is optimized following a structural constraint.
This experiment represents a first step towards a fully autonomous smart network, where communication rates are automatically optimized based on bidirectional information exchange between the users and the network provider.
We consider a fully-connected entanglement network where one of the user nodes is located far apart from the entangled photon source, while all the other nodes are located close to the source as schematically depicted in Fig.\:\ref{fig:flexgrid}\:(a). We implement this situation for $N=5$ by separating node B's detector from the source by a \SI{25}{\kilo\metre}-long SMF28 fiber spool. If all two-user links have the same fixed bandwidth, the coincidence rate of the four links connecting the distant user to the rest of the network, \emph{i.e.}, AB CB DB EB, will be lower than the others. To avoid this problem, we reallocate the state bandwidth by changing the channel widths, assigning more bandwidth to long links, and less bandwidth to short links, as schematically depicted in Fig.\:\ref{fig:flexgrid}\:(b). We use a simple algorithm and distribute the available bandwidth with a \SI{12.5}{\giga\hertz} resolution corresponding to current ITU standards for channel bandwidth granularity.

In order to demonstrate that we can efficiently level the measured signal across the whole network, we measure one by one the 10 two-user coincidence counts rates for each link. The results are shown in Fig.\:\ref{fig:flexgrid}\:(c). Red squares represent the raw coincidence counts of the 10 two-user links for fixed \SI{100}{\giga\hertz} channels and blue triangles are the raw coincidence counts recorded after bandwidth reallocation. We observe that, starting from a very unbalanced initial configuration, we can bring all users to a similar level of signal. The same technique could be applied to other ends, such as boosting the signal across specific links according to user needs. This proof-of-principle experiment shows that our scheme is fully compatible with state-of-the-art telecom network management, opening the way for flexible metropolitan-scale entanglement distribution.\\
	

\noindent {\large\textbf{Discussion}}\\
We have demonstrated a scalable approach to a fully-connected entanglement distribution network using an AlGaAs chip emitting broadband polarization-entangled photon pairs in the telecom band. The lower bound on entanglement fidelity of the quantum state generated by our chip stays above \SI{95}{\percent} over a \SI{26}{\nano\metre} wide spectral range around biphoton degeneracy and above \SI{85}{\percent} over a \SI{60}{\nano\metre} range. We deterministically separate the photons of each pair into energy-matched frequency channels and distribute them to the network users using a wavelength selective switch. We benchmark the performance of our quantum network by running the entanglement-based BBM92 QKD protocol. We perform key distribution between two users with a QBER below 2\% across fibered optical links of up to 50 km including finite-key effects and we extrapolate a positive key rate for distances of up to \SI{75}{\kilo\metre} in both symmetric and asymmetric configurations. We have extended our study to the multi-user case, taking advantage of the flexibility offered by our setup. By reconfiguring the frequency grid, we show that our network can accommodate up to 8 users over 50 GHz ITU channels. We further demonstrate the insensitivity of the QBER with respect to channel width, indicating that every two-user link in the network can support high-performance QKD at metropolitan-scale distances. Finally, we demonstrate that the bandwidth reallocation enabled by the WSS can be used to equilibrate unbalanced network scenarii, which is compatible with an elastic network configuration. Future work includes the implementation of our architecture in a metropolitan quantum communication network to test its performance and robustness in a real-world situation.

Further progress is possible in different ways. On one hand, an extreme miniaturization of the AlGaAs source can be achieved thanks to its compliance with electrical injection as already shown in \cite{Boitier2014}, giving a clear advantage to our approach in terms of portability, energy consumption and integration with future quantum technologies. On the other hand, the cavity effects presently limiting the fidelity to a polarization Bell state far from degeneracy can be avoided by applying an anti-reflection coating to the waveguide facets. An optimization of the design can also be implemented to reduce the modal birefringence in order to further broaden the entangled photon pair bandwidth \cite{Chen2019} and to place the degeneracy wavelength at the center of the telecom C-band to fully exploit the WSS bandwidth. Using WSS technology covering both the C and L bands would further allow to exploit the full bandwidth of our source. Besides, the biphoton bandwidth and brightness could be optimized by choosing the optimal sample length taking into account SPDC efficiency and propagation losses.

In addition to the improvement of the photon-pair source, future investigations of different entanglement-distribution topologies can benefit from the flexibility of our setup.
In particular, using the full potential of our \SI{60}{\nano\metre} entanglement bandwidth, non-fully connected networks which can host up to 28 users over 50 GHz ITU channels could readily be implemented. In this topology, the network is broken up into independent fully-connected 4-user subnets which are all interconnected via extra links. Owing to the broad bandwidth of our source, this scheme can be demonstrated without the need to resort to passive BS multiplexing, as suggested in Refs.\:\cite{Joshi2020,Liu2020} thus avoiding the reduction of the signal caused by extra optical losses.

Finally, in another approach we could take advantage of the cavity effects present in our sample to generate and coherently control biphoton frequency combs \cite{Maltese2020}, opening the way to the utilization of frequency for quantum information processing \cite{Lukens2016,Fabre2020} and to the combined exploitation of both polarization and frequency entanglement for future quantum networks.
	

\medskip

\noindent {\large\textbf{Acknowledgments}}\\
This work has been supported by ANR (Agence Nationale de la Recherche) through the QUANTIFY Project (Project No. ANR-19-ASTR-0018-01), the French RENATECH network, and by Paris Île-de-France Région in the framework of DIM SIRTEQ through the LION Project. The authors acknowledge Perola Milman for fruitful discussions and Yves Jaouen for the C+L band tunable filter loan.

\medskip

\noindent {\large\textbf{Author contributions}}\\
F.A. built the setup, processed the sample and carried out the experimental work. All the authors contributed to the conception of the experiment. A.L. performed the epitaxial growth of the AlGaAs sample. The paper was written by F.A., E.D. and S.D. All authors discussed the results and commented on the manuscript. E.D. and S.D. supervised the project.	

\medskip

\noindent {\large \textbf{Methods}}\\
\noindent\textbf{Sample and experimental setup.}
The sample consists of a 6-period Al$_{0.80}$Ga$_{0.20}$As/Al$_{0.25}$Ga$_{0.75}$As Bragg reflector (lower mirror), a \SI{366}{\nano\metre} Al$_{0.45}$Ga$_{0.55}$As core and a 2-period Al$_{0.25}$Ga$_{0.75}$As/Al$_{0.80}$Ga$_{0.20}$ Bragg reflector (upper mirror).
Waveguides are fabricated using wet chemical etching to define ~\SI{5}{\micro\meter} wide and around  ~\SI{800}{\nano\meter} deep ridges ; the waveguide length is \SI{4}{\milli\metre}. The sample is pumped with a tunable CW diode laser (TOPTICA TM Photonics DL pro 780) which is coupled into the waveguide through a microscope objective (NA = 0.95, 63$\times$); light emerging from the opposite end is collected with a second identical microscope objective and sent to a fibre coupler, after filtering out the pump wavelength with a high pass filter. A thermocouple and a Peltier cooler, connected to a PID controller, monitor and keep the waveguide temperature constant at \SI{19.3}{\degreeCelsius}, fixing the wavelength degeneracy of the photon pairs to \SI{1556.55}{\nano\metre}, which corresponds to the center of the ITU \SI{100}{\giga\hertz} channel number 26. The demux/mux stage is realized either with a Wavelength Selective Switch (model Finisar 4000s) or with a Coarse Wavelength Division Multiplexing unit (model FS 78163) followed by a tunable C+L band filter (model Alnair Labs CVF-220-CL) depending on the spectral region of interest. After the analysis and distribution stage photons are detected with superconducting nanowire single photon detectors (SNSPD, Quantum Opus) and temporal correlation measurement are performed with a time-to-digital converter (TDC, quTools).\\

\noindent\textbf{Theory.}
We provide the expression of the quantum state of the source and derive the theoretical fidelity $F$ as a function of channel frequency.

We start from the generic form of the two-photon state generated by collinear Type-II parametric downconversion:
\begin{equation}\label{eq:qstate_generic}
\ket{\psi} = \iint_{-\infty}^{+\infty} \mathrm{d}\omega_1 \mathrm{d}\omega_2 \mathcal{C}(\omega_1, \omega_2)
\ket{\omega_1, H}\ket{\omega_2, V}
\end{equation}
where
$\ket{\omega, \alpha} = a^{\dagger}_{\alpha}(\omega) \ket{\text{vac}}$ denotes the state of the electromagnetic field with one photon in a mode of polarization $\alpha=H,V$ and angular frequency $\omega$.
The complex function $\mathcal{C}(\omega_1, \omega_2)$ is called the joint spectral amplitude (JSA) and it is normalized to unity $\iint \mathrm{d}\omega_1 \mathrm{d}\omega_2 |\mathcal{C}(\omega_1, \omega_2)|^2 = 1 $.
By rewriting the quantum state in the rotated basis: $\omega_+ = \omega_1+\omega_2$ and $\omega_- = \omega_1-\omega_2$, the JSA takes the form \cite{Boucher2015}: $\mathcal{C}(\omega_1, \omega_2) = f_+(\omega_+)f_-(\omega_-)$. The first term of the product corresponds to energy conservation during the downconversion process and the second term is related to momentum conservation (phase matching). The latter can be numerically calculated using the dispersion properties of the guided modes involved in the downconversion process. Cavity effects arising from the nonzero reflectivity of the AlGaAs waveguide facets can be included in the numerically-calculated JSA using a two-mode Airy distribution \cite{Maltese2020}. Since the waveguide is pumped by a narrow-linewidth CW laser, one can approximate $f_+$ to a Dirac delta function centered on the pump laser angular frequency $f_+(\omega_+) = \delta(\omega_+-\omega_p)$ and the resulting generated quantum state is strongly frequency-anticorrelated. In this case, the state can be rewritten:
\begin{equation}\label{eq:qstate_rotated}
\ket{\psi}  = \int_{-\infty}^{+\infty} \mathrm{d}\Omega \Phi(\Omega) \ket{\omega_d + \Omega, H} \ket{\omega_d - \Omega, V},
\end{equation}
where we defined for convenience $\omega_d = \omega_p/2$ the degeneracy angular frequency, $\Omega = \omega_-/2$ the detuning with respect to the degeneracy and $\Phi(\Omega) = f_-(\omega_-)$.

We then rewrite the state as a continuous superposition of bipartite polarization-entangled states.
To do so, we split the summation in Eq. \eqref{eq:qstate_rotated} into two parts using the identity: $\int_{\infty}^{\infty} = \int_{0}^{\infty} - \int_{0}^{-\infty}$ then make the change of variable $\Omega\rightarrow-\Omega$ into the second term and finally recombine the two integrals to obtain:
\begin{equation}\label{eq:qstate_superposition_bell}
\begin{aligned}
\ket{\psi} = \int_{0}^{\infty} \mathrm{d}\Omega & \left[\Phi(\Omega) \ket{\omega_d + \Omega, H} \ket{\omega_d - \Omega, V} \right.\\
+&\left. \Phi(-\Omega) \ket{\omega_d - \Omega, H} \ket{\omega_d + \Omega, V} \right].
\end{aligned}
\end{equation}

To calculate the fidelity to a $\ket{\Psi^+}$ Bell state as a function of channel frequency, we derive the reduced density matrix in polarization space after spectral filtering. At the demultiplexing stage, signal and idler photons lying in energy matched frequency channels are sent onto separate fiber paths $A$ and $B$. Namely,
all signal photons within the frequency window $[\omega_d + (\Omega_0-\Delta/2), \omega_d + (\Omega_0+\Delta/2)]$ are sent in path $A$ and all idler photons within the frequency window $[\omega_d - (\Omega_0+\Delta/2), \omega_d - (\Omega_0-\Delta/2)]$ in path $B$, where $\Delta$ is the channel width and $\Omega_0$ the detuning of the channel central frequency with respect to degeneracy. The resulting post-selected quantum state takes the form:
\begin{equation}\label{eq:qstate_postselected}
\begin{aligned}
\ket{\psi'} =  \int_0^\infty &\mathrm{d}\Omega f(\Omega)  \left[\Phi(\Omega) \ket{\omega_d + \Omega, H}_A \ket{\omega_d - \Omega, V}_B \right.\\
&+ \left.\Phi(-\Omega) \ket{\omega_d - \Omega, H}_B \ket{\omega_d + \Omega, V}_A \right]
\end{aligned}
\end{equation}
with $f(\Omega)$ the filter lineshape, which in our case is assumed to be rectangular:
\begin{equation}\label{eq:filter}
f(\Omega) = \left\{
\begin{array}{ll}
1,\ &\text{for}\ \Omega \in [\Omega_0-\Delta/2, \Omega_0+\Delta/2] \\
0,\ &\text{elsewhere}.
\end{array}
\right.
\end{equation}
The corresponding density operator is $\tilde{\rho} = \ket{\psi'}\bra{\psi'}$. By following the approach of Ref.~\cite{Schlager2017}, we compute the reduced polarization density matrix $\rho$ by tracing out the frequency part of the density operator:
\begin{equation}\label{eq:densitymatrix_trace}
\rho = \frac{1}{\mathcal{N}} \iint \mathrm{d}\omega' \mathrm{d}\omega'' _A\bra{\omega'}_B\bra{\omega''} \tilde{\rho} \ket{\omega''}_B \ket{\omega'}_A.
\end{equation}
with $\mathcal{N}$ a normalization constant. After some straightforward algebra, one obtains:
\begin{equation}\label{eq:densitymatrix_polar}
\begin{aligned}
\rho &= \alpha \ket{HV}_{ABAB}\bra{HV} + \mathcal{D} \ket{HV}_{ABAB}\bra{VH} \\
& + \mathcal{D^*} \ket{VH}_ {ABAB}\bra{HV} + \beta \ket{VH}_{ABAB}\bra{VH},
\end{aligned}
\end{equation}
where the 4 non-zero matrix elements are:
\begin{align}
\alpha &= \frac{1}{\mathcal{N}} \int_0^\infty \mathrm{d}\Omega f(\Omega) |\Phi(\Omega)|^2 \\
\beta &= \frac{1}{\mathcal{N}} \int_0^\infty \mathrm{d}\Omega f(\Omega) |\Phi(-\Omega)|^2 \\
\mathcal{D} &= \frac{1}{\mathcal{N}} \int_0^\infty \mathrm{d}\Omega f(\Omega) \Phi(\Omega) \Phi^*(-\Omega)
\end{align}
and the normalization constant is set to $\mathcal{N} = \int_0^\infty \mathrm{d}\Omega f(\Omega) \left[|\Phi(\Omega)|^2 + |\Phi(-\Omega)|^2\right]$ such that $\Tr \rho = 1$.
Finally, the fidelity to a $\ket{\Psi^+}$ Bell state can be evaluated from the definition: $F = \left(\Tr \sqrt{\sqrt{\rho} \ket{\Psi^+}\bra{\Psi^+} \sqrt{\rho}}\right)^2$.\\
	
\noindent\textbf{Coincidence histograms and Bell correlation curves.}
We display an example of data for the 8 projective measurements that are used to estimate the Bell state fidelity and the corresponding Bell correlation curves.

\begin{figure*}
	\includegraphics[width=\textwidth, keepaspectratio]{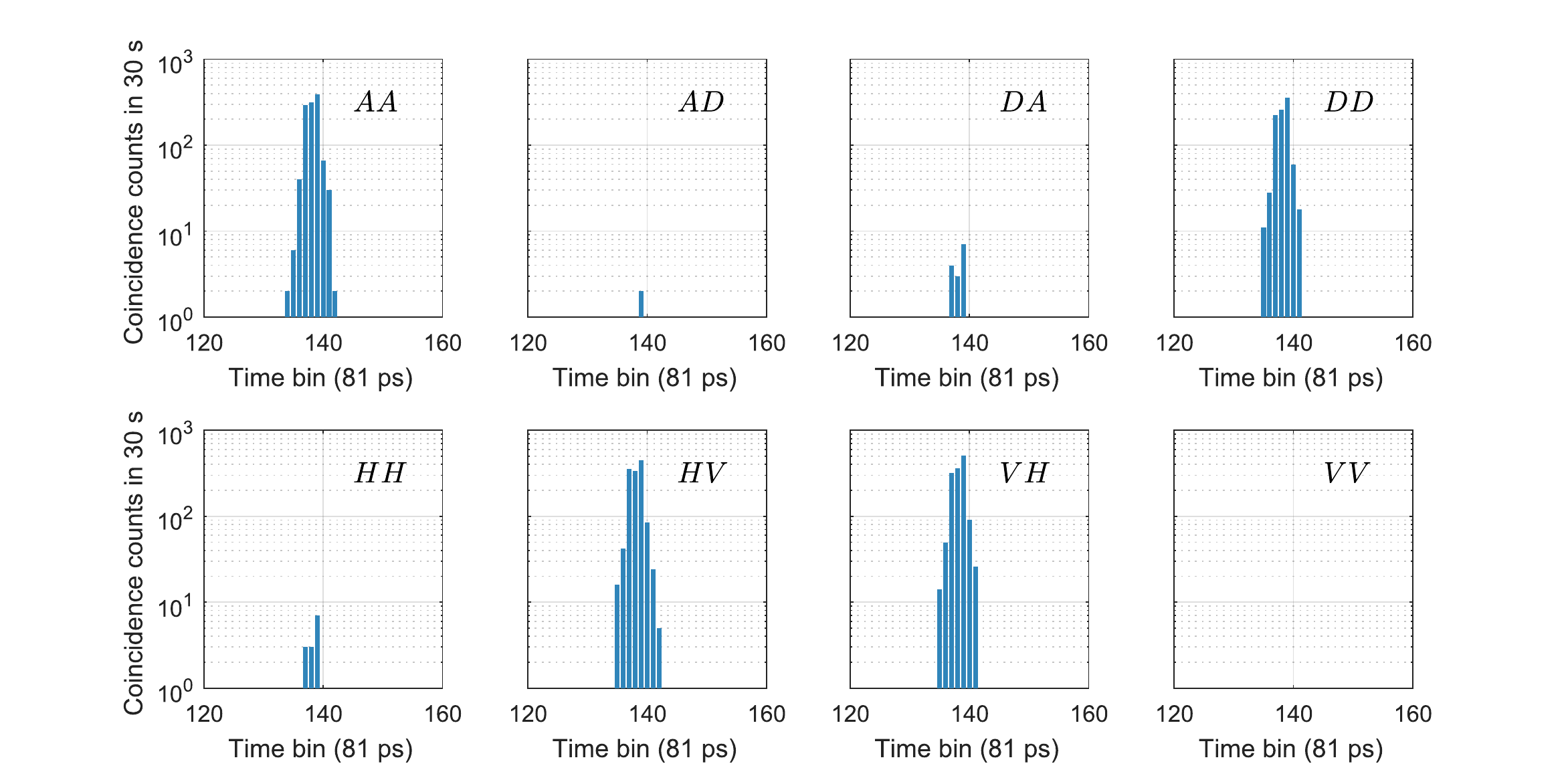}
	\caption{\label{fig:histograms} Measured coincidence histograms for the 8 projective measurements performed to obtain the lower bound on $F$ for 100 GHz ITU channels 23 (\SI{1558.98}{\nano\metre}) and 29 (\SI{1554.13}{\nano\metre}).}
\end{figure*}

The time-correlation histograms that have been recorded for \SI{100}{\giga\hertz} ITU channels 23 and 29 are displayed in Fig.\ref{fig:histograms}. The number of coincidence counts in each configuration is given by the sum of the of the counts in the 6 central time-bins. The width of the time bins is \SI{81}{\pico\second} corresponding to the temporal resolution of our TDC.

We define the coincidence to accidental ratio (CAR) as the mean number of counts in the 6 central bins over the mean number of counts in 6 bins outside the coincidence window. In the configurations where the number of counts is maximum ($AA, DD, HV, VH$) we achieve a CAR of the order of \num{4e3}.

\begin{figure}
	\includegraphics[width=0.45\textwidth,keepaspectratio]{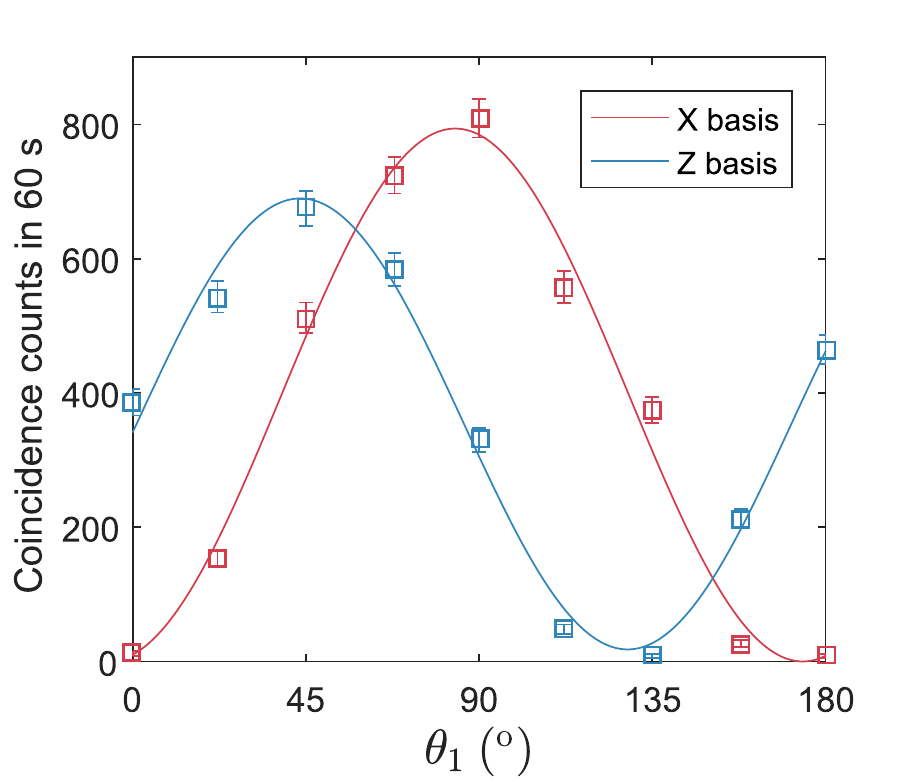}
	\caption{\label{fig:visibility} Measured correlation curves for 100 GHz ITU channels 23 and 29. Squares correspond to experimental raw counts and solid lines to a fit to a sine square.}
\end{figure}
In addition, for each pair of 100 GHz channel, we measured the Bell correlation curves in both $\mathbb{X}$ and $\mathbb{Z}$ bases. To do so, we recorded coincidence counts when projecting the polarization of the signal photon onto an axis of angle $\theta_1$ with respect to the horizontal axis of the laboratory frame, for values of $\theta_1$ spanning $[\ang{0},\ang{180}]$, and the polarization of the idler photon on a fixed axis, either $H\: (\ang{0})$ or $D\: (\ang{45})$. The visibility of the obtained two-photon interference fringes is an indicator of the quality of entanglement. As an illustration, we show the correlation curves obtained between ITU channels 23 and 29 in Fig.\:\ref{fig:visibility}. Solid lines are least-square fits to the expression $\mathcal{C} = a\sin^2(\theta_1 - \theta_2)+b$ where $\theta_1$ and $\theta_2$ are the angles between the horizontal axis of the laboratory frame and the projection axes of signal and idler polarization respectively. We measured raw visibilities of \SI{98.0}{\percent} in the $\mathbb{X}$ basis and of \SI{97.7}{\percent} in the $\mathbb{Z}$ basis.

\bibliography{biblio_qkd}
	
\end{document}